\documentclass[12pt, draftclsnofoot, onecolumn]{IEEEtran}
\usepackage{color}

\ifCLASSINFOpdf
  \usepackage[pdftex]{graphicx}
\else
  \usepackage[dvips]{graphicx}
\fi
\usepackage[cmex10]{amsmath}
\usepackage{amssymb}
\usepackage{amsthm}

\usepackage{array}
\usepackage[round, sort, numbers]{natbib}
\usepackage{booktabs}
\usepackage{multirow}
\usepackage{rotating}
\usepackage{graphicx}
 \usepackage{latexsym}
\usepackage{amsmath}
\usepackage{mdwmath}
\usepackage{mdwtab}
\usepackage{eqparbox}
\usepackage{amsmath,epsfig,psfrag,latexsym,wasysym,amssymb}
\usepackage{varioref}	
\usepackage{amssymb}
\usepackage{color}
\usepackage{epstopdf} 
\usepackage{graphicx, subfigure}
\usepackage{amsthm}
\usepackage{natbib}
\usepackage{textcomp}

\ifCLASSOPTIONcompsoc
  \usepackage[caption=false,font=normalsize,labelfont=sf,textfont=sf]{subfig}
\else
  \usepackage[caption=false,font=footnotesize]{subfig}
\fi

\setcitestyle{square}

\newtheorem{lem}{Lemma}
\begin{document}
%
\title{Precoding in Multigateway Multibeam Satellite Systems}
%
%
%

\author{Vahid Joroughi,
        Miguel \'Angel V\'azquez
				and~Ana I. P\'erez-Neira,~\IEEEmembership{Senior Member,~IEEE}
\thanks{The research leading to these results has received funding from the Spanish Ministry of Science and Innovation under projects TEC2011-29006-C03-02 (GRE3N-LINK-MAC) and the Catalan Government (2014 SGR 1567).}
\thanks{V. Joroughi and A. P\'erez-Neira are with the Universitat Politecnica de Catalunya (UPC) and Centre Tecnol\`ogic de les Telecomunicacions de 
Catalunya (CTTC), Barcelona, Spain.}
\thanks{Emails:vahid.joroughi@upc.edu,~ana.isabel.perez@upc.edu}
\thanks{M. \'A. V\'azquez is with the Centre Tecnol\`ogic de les Telecomunicacions de 
Catalunya (CTTC), Barcelona, Spain.}
\thanks{Emails:mavazquez@cttc.es}}

\maketitle
\begin{abstract}
This paper considers a multigateway multibeam satellite system with multiple feeds per beam. In these systems, each gateway serves a set of beams (cluster) so that the overall data traffic is generated at different geographical areas. Full frequency reuse among beams is considered so that interference mitigation techniques are mandatory. Precisely, this paper aims at designing the precoding scheme which, in contrast to single gateway schemes, entails two main challenges. First, the precoding matrix shall be separated into feed groups assigned to each gateway. Second, complete channel state information (CSI) is required at each gateway, leading to a large  communication overhead.  In order to solve these problems, a design based on a regularized singular value block decomposition of the channel matrix is presented so that both inter-cluster (i.e. beams of different clusters) and intra-cluster (i.e. beams of the same cluster) interference is minimized. In addition, different gateway cooperative schemes are analysed in order to keep the inter-gateway communication low. Furthermore, the impact of the feeder link interference (i.e. interference between different feeder links) is analysed and it is shown both numerically and analytically that the system performance is reduced severally whenever this interference occurs even though precoding reverts this additional interference. Finally, numerical simulations are shown considering the latest fixed broadband communication standard DVB-S2X so that the quantized feedback effect is evaluated. The proposed precoding technique results to achieve a performance close to the  single gateway operation even when the cooperation among gateways is low.
\end{abstract}
\begin{IEEEkeywords}
Multibeam satellite systems, multigateway transmission, MIMO precoding.
\end{IEEEkeywords}
%
\IEEEpeerreviewmaketitle
\section{Introduction}
\IEEEPARstart{W}{henever} multibeam satellite systems increase their overall throughput, their feeder link (i.e. the communication link between the gateway and the satellite) communication shall increase proportionally. Indeed, since the feeder link aggregates all satellite traffic, the resulting required bandwidth is
\begin{equation}
 B_{\text{feeder-link}} = NB_{\text{user-link}},
\end{equation}
where $N$ is the number of feed signals, $B_{\text{user-link}}$ the per-beam bandwidth and $ B_{\text{feeder-link}}$ the feeder link required bandwidth. Considering that the number of feed signals is proportional to the number of beams ($K$) and 
\begin{equation}
N > K,
\end{equation}
it is evident that if the user available bandwidth is increased, the feeder link resources must be increased accordingly. 

Augmenting the user available bandwidth can be performed if an aggressive frequency reuse among beams is deployed. In that case, interference mitigation techniques become mandatory as adjacent beams suffer from a large multiuser interference. Similar reasoning occurs in cellular terrestrial systems when spectrum sharing is considered \cite{jorswieck14}.

Interference mitigation techniques are divided into two main approaches. One approach is that the receiver detects and, posteriorly, subtracts the multiuser interference. Examples of these advanced receivers can be found in \cite{andrea14,amina11}. Another approach is to revert the interference effect at the transmit side, leading to high reduction of the receiver complexity since single user detection can be performed. Under this context, the transmitter must precode the feed signals in order to mitigate in advance the multiuser interference \cite{dev11}. However, precoding is sensitive to  channel state information (CSI) quality which is difficult to obtain due to the feedback and quantization errors. In any case, deciding where to implement the interference mitigation techniques exceeds the scope of this paper and in the sequel we will focus on precoding techniques.

Apart from the already mentioned limitation of precoding, multibeam satellite systems have also several technical challenges \cite{Christopoulos682142}. As mentioned at the beginning, one of them is the feeder link bandwidth requirements. In \cite{bhavani2014} the authors propose a feeder link system operating at the Q/V, where there is a larger available bandwidth than in the Ka band  (currently used for this service). Although this frequency band can support the required feeder link bandwidth, rain attenuation severally decreases the performance with respect to the Ka band. Another option is to use multiple gateways and reuse among them the Ka band available bandwidth. As a result, the gateways must support the feeder links with very directive antennas so that the bandwidth can be fully reused among them. In any case, whenever multiple gateways are employed, precoding techniques shall be reconsidered \cite{zheng2012,asilomar}.

First, since the data traffic is independently generated at each gateway, every gateway must precode the signal in a decentralized fashion and transmit through their corresponding feeder link. In other words, the overall precoding matrix is distributively computed at each gateway so that each gateway can only use certain feed elements. Second, since for computing the precoding matrix each gateway requires CSI from other gateways, certain inter-gateway communication is required, leading to large system resources overhead.

This paper deals with the problem of precoding in multigateway multibeam satellite systems. In contrast to previous works, this paper studies the case where the collaboration between different gateways is limited. In \cite{zheng12} a power optimization scheme is presented assuming linear mean square error (LMMSE) precoding, full information sharing among gateways and  single feed per beam architecture. On the contrary, our proposal considers the multiple feed per beam structure which is a more general payload architecture and we study the case where the information exchanged between gateways is reduced. 

Other works regarding multiple gateway precoding are \cite{asilomar, ew}. In \cite{asilomar} a first approach to the problem is presented considering that precoding was able to block completely the interference between gateways which is not true in general due to the limited number of feed signals. Furthermore, in \cite{ew} a modification of the technique presented in \cite{asilomar} is described considering per beam power constraints.

In addition to the aforementioned contributions, in this paper some unexplored aspects of multigateway multibeam precoding are investigated. First, the tentative impact of feeder link interference is considered. Note that this effect occurs whenever the payload feeder link receiver is not properly calibrated and certain pointing errors take place. Second, the limited cooperation among gateways is considered and a matrix compression technique is provided which shows a good trade off between increased overhead and overall throughput. Finally, the impact of imperfect CSI is evaluated considering the current standard  digital video broadband (DVB-S2X).

To summarize, the contributions of this paper are:
\begin{itemize}
\item A precoding technique for multiple gateway architecture where the transmit data and CSI is distributed among gateways.
\item The feeder link interference is studied and the mean square error (MSE) is characterized under this effect.
\item A proposal for reducing the communication overhead between gateways is presented.
\item The effect of limited feedback from the users and among gateways is evaluated assuming the latest standard for broadband satellite communications.
\end{itemize}

%
%
%
%

The rest of the paper is organized as follows. Section II describes the multigateway multibeam system architecture so as its signal model. Section III provides a novel multigateway precoding technique for multiple feed per beam payloads. In section IV is described the limitations of multigateway multibeam precoding techniques. Section V shows the numerical simulations and Section VI concludes.

\textbf{Notation}: Throughout this paper, the following notations will be adopted. Boldface uppercase letters denote matrices and boldface lowercase letters refer to column vectors. $ (.)^H$ and $(.)^T $   denote  Hermitian transpose and transpose   matrices, respectively.  $\mathbf{I}_N$ builds $ N\times N$ identity matrix. Boldface  $\mathbf{0}$ and $\mathbf{1}$ refer to an all-zero matrix and  the vector of ones, respectively.  $\mathbf{A}_{ij}$ represents the ($i$th, $j$th) element of matrix $\mathbf{A}$ and $(\mathbf{A})_{K\times K}$ denotes a submatrix of $\mathbf{A}$ of size $K \times K$.  diag represents a diagonal matrix. $\text{Tr}$ denotes the trace operator.  Finally, $\mathrm E \{\mathbf{.}\}$ and $||.||$ refer to the expected value  operator and the Frobenius norm, respectively.

\section{System Description}

Consider a multibeam satellite communication system where a single geosynchronous satellite with multiple beams provides broadband services to a large set of users.  The coverage area is divided into $K$ beams and the users are assumed to be uniformly distributed within each beam.

By employing a Time Division Multiplexed (TDM) scheme, at each time instant a total of $K$ single antenna users, i.e. exactly one user per beam, is simultaneously served by multiple gateways. The total number of gateways is assumed to be $G$. In this context, each gateway serves a set of adjacent beams and the satellite is equipped with $G$ feeder link receivers, each one associated to a single gateway. Although in this first system model definition we do not include a tentative interference between feeder links, in subsequent sections we will investigate its possible impact. In any case, we will assume that the feeder link communication is ideal (i.e. noiseless communication) which is a realistic assumption considering the link budget of these systems.

By assuming full frequency reuse pattern among beams,  the received signals can be modelled as
\begin{equation}
\mathbf{y}=\mathbf{H}\mathbf{x}+\mathbf{n},
\label{signal model}
\end{equation}
where  $\mathbf{y}\in \mathbb{C}^{K \times 1}$ is a vector that contains the symbols received by  each user, $\mathbf{x} \in \mathbb{C}^{N \times 1}$ denotes the stack of the transmitted signals at all feeds and $\mathbf{n} \in \mathbb{C}^{K \times 1}$ contains the stack of zero mean unit variance Additive White Gaussian Noise (AWGN) such that $\mathrm E \{\mathbf{n}\mathbf{n}^H\}=\mathbf{I}_K$. Finally, $\mathbf{H} \in \mathbb{C}^{K \times N}$ is the overall user link  channel matrix whose element $(\mathbf{H})_{i,j}$ presents the  gain of the link between the i-th user (in the i-th beam) and the j-th satellite feed.  
Focusing on the channel model, the user link channel matrix $\mathbf{H}$ can be decomposed as follows:
\begin{equation} 
\mathbf{H} =\mathbf{DW},
\label{channel dec}
\end{equation}
where:
\begin{itemize}
\item $\mathbf{D}$ is assumed to be a $K\times K$ diagonal matrix which takes into account the atmospheric fading in the user link such that
\begin{equation}
\mathbf{D}=\text{diag}\left(\frac{1}{\sqrt{A}_1}, ..., \frac{1}{\sqrt{A}_K}\right),
\end{equation}
where $A_k$ denotes the rain attenuation affecting the $k$-th beam.
 \item $\mathbf{W}$ is a $K\times N$ matrix which models the feed radiation pattern, the path loss, the receive antenna gain and the noise power. The $(k,n)$-th entry of $\mathbf{W}$ is modelled as
\begin{equation}
w_{kn}=\frac{W_R\  g_{kn}}{4\pi \frac{d_k}{\lambda}\sqrt{k_B T_R
BW}},
\end{equation}
where $W_R^2$ denotes the user  receive antenna's power gain. $g_{kn}$ is a complex value which models the radiation pattern from from the $n$-th to the $k$-th user, such that the respective feed transmit gain  is $10\log_{10}(|g_{kn}|^2)$ if expressed in dBi. Finally, $d_k$  is the distance between the $k$-th user  and the satellite, $\lambda$ the carrier wavelength, $k_B$ the Boltzmann constant, $T_R$ the receiver noise temperature, $BW$ the carrier bandwidth. Note that the elements of $\mathbf{W}$ are normalized by unit  variance noise. 
\end{itemize}

The reader can refer to \cite{miguel11} for a more detailed description of the user link channel model. Assuming that the symbols are linearly processed before being transmitted, the transmit data can be written as
\begin{equation}
\mathbf{x}= \mathbf{Ts},
\label{precoder}
\end{equation}
where $\mathbf{s} \in \mathbb{C}^{K \times 1}$ is the transmit symbol vector such that the $k$-th entry of $\mathbf{s}$ is the constellation symbol destined to the $k$-th user with  $\mathrm E \{\mathbf{ss}^H\}=\mathbf{I}_K$. $\mathbf{T} \in \mathbb{C}^{N \times K}$ denotes the block linear precoding matrix. Remarkably, $\mathbf{T}$ is computed at the gateways and it shall be transmitted through the feeder links. Thus, in the feeder link transmission the precoding matrix shall be separated in blocks and simultaneously transmitted by the gateways. This will be presented in the following sections. 

The typical performance metric involves the received signal to noise plus interference ratio (SINR) for $k$-{th} user, which is expressed as
\begin{equation}
\text{SINR}_{k}=\frac{|({\mathbf{H}}\mathbf{T})_{kk}|^2}{\sum_{j\neq k}^{K}|({\mathbf{H}}\mathbf{T})_{kj}|^2+1} \quad k=1, \ldots, K.
\label{SINR}
\end{equation}

To generate a power flexibility  which is essential for optimum resource allocation in multibeam system, we assume on-board travelling-wave tube amplifiers (TWTAs). In general, in TWTAs symbols are perfectly oversampled and pulse-shaped with a square-root raised cosine (SRRC) filter and a small roll-off factor. Therefore, these  high power amplifiers provide a large bandwidth and high signal power level with small level of interference among feeds' signals. In fact, the total power obtained by  a set of TWTAs can be  distributed among different feeds and the number of TWTAs is related to the number of feeds used in the antenna. Under this context, multiple feeds will distribute the available power to each beam. As a result, the precoding design is based on total power constraint. Mathematically,
\begin{equation}
 \mathrm E \{||\mathbf{x}||^{2}\}= \text{trace}(\mathbf{T}\mathbf{T}^{H}) \leq P,
\label{p_dist}
\end{equation}
where $P$ denotes the total transmit power at the satellite.

\begin{figure}
\centering
\includegraphics[scale=0.5]{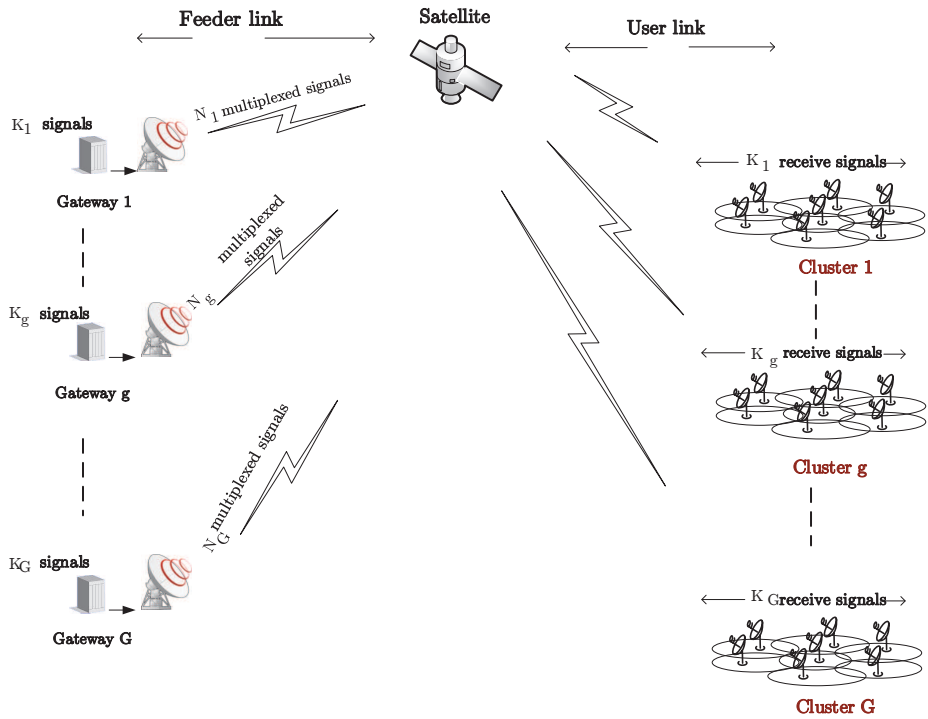}\caption{ Multiple gateway structure. The transmitted symbols are produced in geographically separated areas and they are transmitted separately through the feed signals. The satellite is equipped with an array fed reflector antenna where those $N = \sum_{g=1}^G N_g$ feed signals are transformed into $K$ transmitted user signals. Interference is created not only by the signals within each cluster but also within the different clusters. The number of beams (users) per cluster is equal to $K_g$.}
 \label{fig:m_GW}
\end{figure}

This paper is devoted to the study of multiple gateway precoding transmission so that in the following we will assume that beams are divided into clusters and each cluster is served by a certain gateway. There is the same number of gateways and clusters as Figure 1 depicts. As we remarked previously, each feed signal can only be generated by a single gateway, otherwise signal overlapping might occur. Considering that there are $G$ gateways, we will denote $N_g$ the number of feeds associated to the $g$-th gateway for $g =1, \ldots , G$. Note that
\begin{equation}
N = \sum_{g=1}^G N_g.
\end{equation}
In addition, the number of served beams by the $g$-th gateway is defined as $K_g$ for $g=1, \ldots , G$ where
\begin{equation}
K = \sum_{g=1}^G K_g.
\end{equation}
In addition, we will consider that 
\begin{equation}
K_g \leq N_g \quad g=1, \ldots , G.
\end{equation}
It is important to remark that the optimization of $K_g$ and $N_g$ for $g=1, \ldots G$ is out of the scope of this paper and we will assume beforehand a certain feed and beam allocation per gateway. Consequently, each gateway must compute a precoding matrix $\mathbf{T}_g \in \mathbb{C}^{N_g \times K_g}$. In other words, each gateway must compute the linear processing for serving $K_g$ users with $N_g$ feeds. Figure \ref{fig:m_GW} shows the overall system architecture.

\begin{figure}
\centering
\includegraphics[scale=0.3]{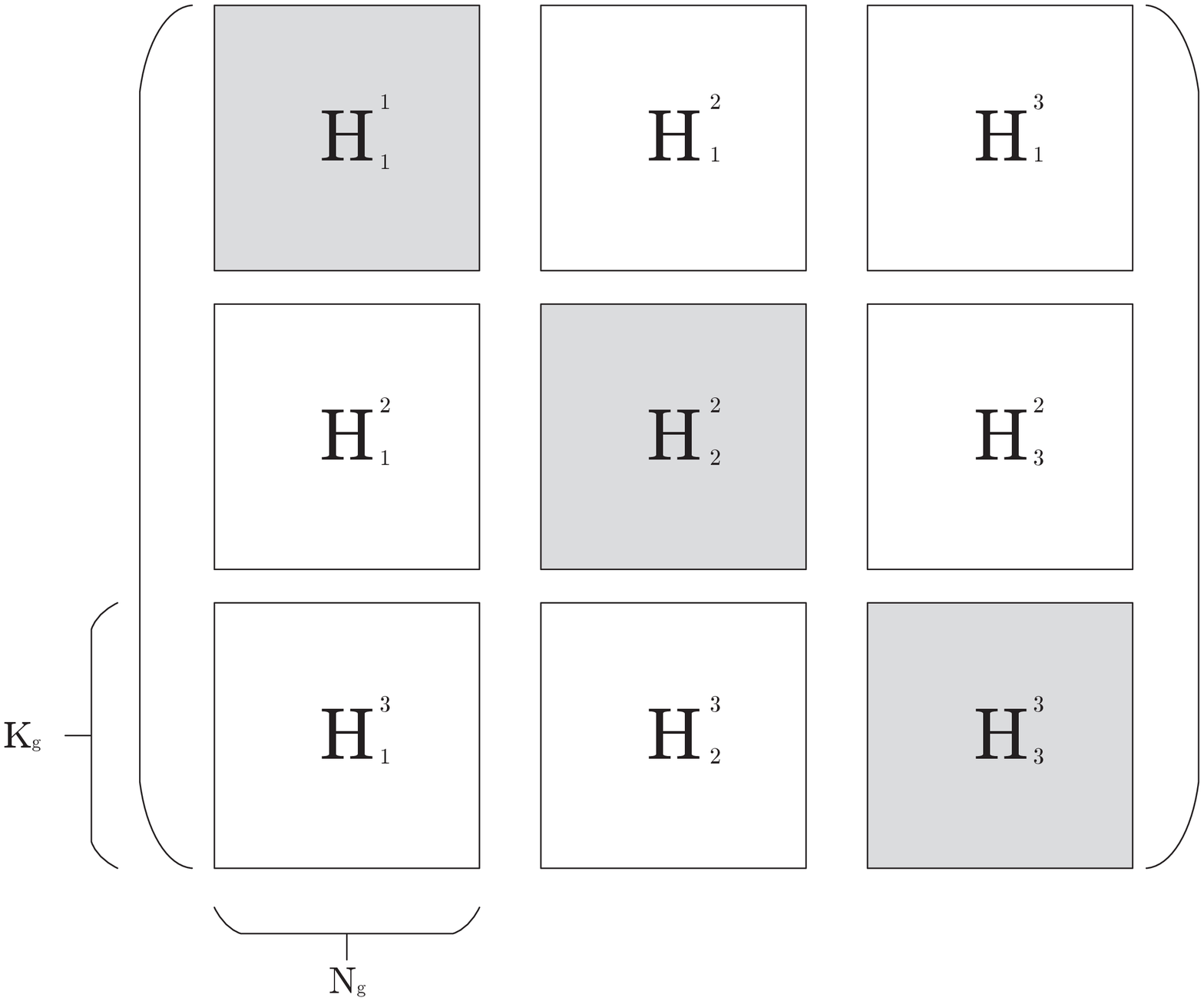}
\caption{ Representation an example of the channel matrix of a multibeam multigateway satellite system with three gateways. The grey sub-matrices represent the channel effect of the feed signals assigned to a certain gateway to the users assigned to the gateway. The white sub-matrices represent the impact of the feed signals to the non-intended clusters. The role of the precoding matrix is not only to revert the interference of the users within the same cluster (grey sub-matrix) but also to mitigate the interference generated to the rest (white sub-matrices).}
\label{matrix}
\end{figure}

Prior to designing the precoding matrix, let us define the following division of the channel matrix
\begin{equation}
\mathbf{H} = \left(\mathbf{H}_1, \ldots, \mathbf{H}_G \right),
\end{equation}
where $\mathbf{H}_g \in \mathbb{C}^{K \times N_g}$ is the channel sub-matrix containing the contribution of the feeds assigned to the $g$-th gateway. Note that we will assume that the allocated feed elements for each gateway are consecutive in the channel matrix. Moreover, each sub-matrix can be decomposed into clusters as
\begin{equation}
\mathbf{H}_g = \left(\mathbf{H}_g^{1,H}, \ldots,  \mathbf{H}_g^{G,H}\right)^H,
\end{equation}
where $\mathbf{H}_g^{c} \mathbb{C}^{K_c \times N_g}$ for $c=1, \ldots , G$ is the channel sub-matrix corresponding to the effect of the $N_g$ feeds to the $c$-th cluster. An illustrative example of this sub-matrix decomposition is represented in figure \ref{matrix}.

Next sections are devoted to the design of $\mathbf{T}_g$ for $g=1, \ldots , G$ in order not only to mitigate the interference between clusters but also the interference created within each cluster. Posteriorly, limitations of this precoding technique are studied.

\section{Multigateway Multibeam Precoding}

As it is described in the previous section, the multiple gateway precoding does not only have to reject the intra-cluster interference but also the inter-cluster one. For the latter, it is proposed a block regularized singular value decomposition (SVD) precoding. Considering the $g$-th gateway channel submatrix, let us construct its regularized version
\begin{equation}
\mathbf{H}_g^{(R)} = \mathbf{H}_g\mathbf{H}_g^H + \frac{G}{P}\mathbf{I}_K \in \mathbb{C}^{K \times K},
\end{equation}
which is used to construct the precoder instead of using $\mathbf{H}_g$. The reason is that the null space has zero dimension and, as a result, it is impossible to block the inter-cluster interference. Nevertheless, as it will be shown in the simulation section, the out-of-cluster interference is mitigated severally if the regularized version is used instead. This technique was also applied in \cite{silva09} for the multigroup multicast scenario. To understand how, let us  decompose $\mathbf{H}_g^{(R)}$, in the following sub-matrices
\begin{equation}
\mathbf{H}_g^{(R)} = \left(\mathbf{H}_g^{(1,R),H}, \ldots, \mathbf{H}_g^{(g,R),H}, \ldots,   \mathbf{H}_g^{(G,R)}\right)^H,
\end{equation}
where $\mathbf{H}_g^{(g,R),H}\in \mathbb{C}^{ K_g \times K}$ is the sub-channel matrix of the $g$-th cluster in the regularized matrix. In addition, the following matrix can be constructed with the out-of-cluster sub-matrices
\begin{equation}
\widetilde{\mathbf{H}_g^{(g,R)}} = \left(\mathbf{H}_g^{(1,R),H}, \ldots, \mathbf{H}_g^{(g-1,R),H}, \mathbf{H}_g^{(g+1,R),H}, \ldots,   \mathbf{H}_g^{(G,R),H}\right)^H .
\end{equation}
Considering the singular value decomposition of this last matrix,
\begin{equation}
\widetilde{\mathbf{H}_g^{(g,R)}} = \mathbf{U}_{g}\boldsymbol{\Lambda}_{g}\mathbf{V}_{g}^H,
\end{equation}
where $\mathbf{V}_{g}$ can be written as
\begin{equation}
\mathbf{V}_{g} = \left(\mathbf{V}_{g}^1 \mathbf{V}_{g}^0 \right)^H,
\end{equation}
where $\mathbf{V}_{g}^1 \in \mathbb{C}^{K \times (K-K_g)}$ are the left singular eigenvectors of $\widetilde{\mathbf{H}_g^{(g,R)}}$ associated to the non-zero singular eigenvalues, whereas $\mathbf{V}_{g}^0\in \mathbb{C}^{K \times (K_g)}$ are the left singular eigenvectors associated to the zero singular eigenvalues. In this context, the gateway can use $\mathbf{V}_{g}^0$ as a pre-processing in order to mitigate the interference created by its feeds to the users outside its corresponding cluster. 

Now it is time to design the precoding matrix in order to mitigate the intra-cluster interference. Considering that each gateway employs the aforementioned pre-processing matrix, the gateway observes a virtual channel matrix 
\begin{equation}
\mathbf{H}_g^{\text{eq}} = \mathbf{H}_g^{(g,R)}\mathbf{V}_{g}^0 \in \mathbb{C}^{K_g \times K_g}.
\end{equation}
With this, the system designer can resort to the common designs such as zero forcing (ZF),
\begin{equation}
\mathbf{W}_{g, \text{ZF}}= \mathbf{H}_g^{\text{eq},H}\left(\mathbf{H}_g^{\text{eq}} \mathbf{H}_g^{\text{eq},H} \right)^{-1}.
\label{madar2}
\end{equation}
Moreover, an additional gain can be obtained by means of employing the regularized zero forcing precoding or linear minimum mean square error (LMMSE)
\begin{equation}
\mathbf{W}_{g, \text{LMMSE}}= \mathbf{H}_g^{\text{eq},H}\left(\mathbf{H}_g^{\text{eq}} \mathbf{H}_g^{\text{eq},H} + \frac{P}{G}\mathbf{I}_{K_g} \right)^{-1}.
\label{madar3}
\end{equation}
The final precoder is computed so that
\begin{equation}
\mathbf{T}_{g} = \beta\mathbf{H}_g^{c,H}\mathbf{V}_{g}^0\mathbf{W}_{g} \in \mathbb{C}^{N_g \times K_g},
\end{equation}
for both the ZF and MMSE case and where $\beta$ is set so that
\begin{equation}
\text{Tr}\left(\mathbf{T}_{g}^H\mathbf{T}_{g}\right) = \frac{P}{G} \quad g=1, \ldots , G,
\end{equation}
where it has been assumed that the total power is equally divided for each gateway and the power constraints are fulfilled with equality. Finally, constructing the overall precoding matrix is done via
\begin{equation}
\mathbf{T} = \text{block-diagonal}  \{  \mathbf{T}_{g} \}_{g=1}^G,
\end{equation}
where the block-diagonal operator constructs a matrix formed by the inputs matrices in the main diagonal and zero otherwise.

To sum up, the multigateway design consists of two stages. First, a pre-processing matrix that minimizes the inter-cluster interference is presented based on a regularized SVD decomposition. Second, for the resulting virtual cluster channel, two precoding techniques are considered: ZF and LMMSE. These latter techniques will mitigate the intra-cluster multiuser interference. Remarkably, the computation of the second precoding matrices can be done separately at each gateway where the first crucial pre-processing stage requires cooperation among gateways.

In the next section precoding limitations are presented. Note that the multigateway precoding relays on the assumption that gateways know the overall channel matrix and; in addition, this CSI has no errors. These two assumptions are generally unrealistic. The impact of limited CSI so as a possible reduction on the cooperation among gateways is studied in the next section. Finally, an analysis of the feeder link calibration errors is shown.

\section{Multigateway Precoding Limitations}

\subsection{Reduced cooperation among gateways}

The regularized block SVD multigateway precoding presented in the previous section assumes that each gateway knows $\mathbf{H}_g$. This assumption entails that each gateway is able to receive information by its feeder link from not only the users of its cluster but also from the rest. Unfortunately, this architecture will require complex data routing mechanisms at the payload which entails an unaffordable satellite hardware complexity.

Assuming that the gateways are connected by a high speed fibre optic and assuming that satellite channel variations are minimal, full CSI sharing among gateways is implementable at expenses of a large communication overhead among gateways. In any case, the system designer might not decide to allocate such an overhead of resources for gateway cooperation and he/she can decide to keep them low. In that case, limited cooperation techniques are required.

Concretely, the $g$-th gateway knowns
\begin{equation}
\{\mathbf{H}_{i}^g \}_{i = 1}^G, \quad g=1, \ldots, G.
\end{equation}
However, each gateway needs to know
\begin{equation}
\mathbf{H}_g, \quad g=1, \ldots, G.
\end{equation}
As a result, each gateway should share 
\begin{equation}
\{\mathbf{H}_{i}^g \}_{i \neq g}^G, \quad g=1, \ldots, G,
\end{equation}
which is a total amount of $N_g(K - K_g)G$ complex numbers to be transmitted through the connection between gateways. In order to reduce this communication overhead, the following approximation is presented. 

Considering the operation at the $g$-th gateway, the CSI information to be transmitted to the $i$-th gateway is
\begin{equation}
\mathbf{H}_{i}^g \in \mathbb{C}^{K_g \times N_g} \quad i\neq g.
\end{equation}
The best rank-one approximation of the aforementioned matrix is given by the right singular eigenvector associated to largest singular eigenvalue: $\sigma_i\mathbf{v}_{g,i} \in \mathbb{C}^{N_g \times 1}$. In this way, the $g$-th gateway can construct a matrix which approximates the inter-cluster interference as
\begin{equation}
\mathbf{G}_g = \left( \sigma_1\mathbf{v}_{g,1}^H, \ldots, \sigma_{g-1}\mathbf{v}_{g,g-1}^H, \sigma_{g+1}\mathbf{v}_{g,g+1}^H, \ldots ,  \sigma_{G}\mathbf{v}_{g,G}^H\right)^H.
\end{equation}
Under this context, each gateway only needs to transmit $N_g(G-1)$ complex numbers instead of the  $N_g(K - K_g)G$ when full cooperation is considered. Matrix $\mathbf{G}_g \in \mathbb{C}^{(G-1) \times N_g}$ is an approximation of $\mathbf{H}_g$. Similar to the previous method, the null space project of $\mathbf{G}_g$ can be used as a pre-processing matrix following the scheme presented in the previous section. This development is not included in the paper for the sake of brevity.

Apart from the presented reduction in terms of the singular value decomposition of $\mathbf{H}_{i}^g$, the system designer can limit the cooperation overhead by means of reducing the communication between gateways. With this, only a subset of gateways interchange CSI data leading to a large reduction of the communication overhead. In the simulation section this is carefully evaluated.

\subsection{Feeder link interference}

So far we have considered that the gateways are able to transmit the feed signals within several feeder links (i.e. one feeder link per gateway) in an interference-free and noiseless channel. Indeed, the satellite is equipped with $G$ feeder link receivers so that the signals are demultiplexed and routed through the array fed reflector. Unfortunately, although very directive antennas are used on ground for transmitting the feed signals, pointing errors might occur and; in addition, hardware might become uncalibrated under certain conditions. Consequently, an additional interference among beams might be created due to this.

Mathematically, the channel matrix becomes
\begin{equation}
\mathbf{H} \leftarrow \mathbf{H}_u\mathbf{H}_{f},
\end{equation}
where $\mathbf{H}_{f} \in \mathbb{C}^{N \times N}$ models the interference generated by the different gateways and $\mathbf{H}_u$ is the channel matrix described in the previous system model section. It is assumed that the interference equally impacts all the feed signals of each gateway (i.e the feeder link channel is not frequency selective). 

In the following we provide a tentative description of $\mathbf{H}_{f}$ matrix. Assuming that $\mathbf{H}_{f}$ is formed by block matrices, $\mathbf{H}_f^{i,j}$ of dimension $N_g \times N_g$ which model the each gateway interference, the feeder link interference impact can be described as
\begin{equation}
\mathbf{H}_f^{i,i} = \mathbf{I}_{N_g},
\end{equation}
\begin{equation}
\mathbf{H}_f^{i,j} = \rho^{|i - j|}\mathbf{E}_{N_g} \quad i \neq j,
\end{equation}
for $i = 1, \ldots , G$ and where $\mathbf{E}_{N_g}$ is a $N_g \times N_g$ whose entries are equal to one. Moreover, $\rho \in [0,1]$ is a parameter that models the overall interference signals. The larger $\rho$ the larger feeder link interference is considered.

Intuitively, whenever the systems presents larger interference prior the precoding effect, the lower achievable rates will be obtained. This reasoning is mathematically proved in the following by means of considering an upper bound of the sum of MSE (SMSE).

Let us consider that the inter-cluster interference is completely suppressed and each gateway performs LMMSE precoding. This scenario is an upper-bound of the overall multibeam multigateway system performance since in general the inter-cluster interference cannot be neglected. Under this context, the forward link system achievable rates under linear precoding considering an arbitrary $g$ cluster can be described by the SMSE \cite{dev11}. Mathematically, 
\begin{equation}
\text{SMSE}=\text{trace}\big(\textbf{MSE}\big).
\label{sum_rate12}
\end{equation}
where $\textbf{MSE}=\text{diag}\big(\text{MSE}_{1}, ..., \text{MSE}_{i}, ..., \text{MSE}_{K} \big)$ and where $\text{MSE}_{i}$ refers to the MSE received by $i$-th user. Obviously, the lower SMSE, the larger achievable rates can be obtained. It can be shown that the SMSE can be written as
\begin{equation}
\text{SMSE} = \frac{K}{P} \text{Tr}\left(\left(\mathbf{I}\frac{K}{P} + \mathbf{H}_u\mathbf{H}_u^H \right)^{-1} \right),
\end{equation}
where the $g$ superscript has been omitted for the sake of clarity. With this last equation, the impact of the feeder link multigateway interference can be analytically studied. Consequently, it is possible to define
\begin{equation}
\text{SMSE}_{\text{no-interference}} = \frac{K}{G} \text{Tr}\left(\left(\frac{G}{P}\mathbf{I} + \mathbf{H}_u\mathbf{H}_u^H \right)^{-1} \right),
\end{equation}
\begin{equation}
\text{SMSE}_{\text{interference}} = \frac{K}{G} \text{Tr} \left( \left(\frac{G}{P} \mathbf{I}  + \mathbf{H}_u\mathbf{H}_{f} \mathbf{H}_{f}^H \mathbf{H}_u^H \right)^{-1} \right),
\end{equation}
where we did not include the superscript in $\mathbf{H}_{f}$ for the sake of clarity but it is important to remark that its dimensions are $N_g \times N$. Prior to establishing the relation between the SMSE of these two cases, the following lemma is introduced.

\begin{lem}
Consider~ $\mathbf{D}= [d_1, ..., d_k]$ be any tall matrix and $\mathbf{D}_r=[d_1, ..., d_r]$, for all $\text{r} = 1, \ldots, k-1$ it holds that 
\label{LEMMASD}
\end{lem}
\begin{equation}
\sigma_{1}(\mathbf{D}_{r+1}) \geq \sigma_{1}(\mathbf{D}_{r}) \geq \sigma_{2}(\mathbf{D}_{r+1}) \geq ... \geq \sigma_{r}(\mathbf{D}_{r+1})\geq \sigma_{r}(\mathbf{D}_{r}) \geq \sigma_{r+1}(\mathbf{D}_{r+1})
\label{gen234}
\end{equation}

\begin{proof}
See Theorem 1 in \cite{thompson72}.
\end{proof}

With this result, the following theorem can be established.

\textbf{Theorem 1} For any semidefinite positive matrix $\mathbf{H}_f$, it results that
\begin{equation}
\text{SMSE}_{\text{interference}} \geq \text{SMSE}_{\text{no-interference}}.
\end{equation}
\begin{proof}
See Appendix A.
\end{proof}

From this last result it is evident that whenever the multigateway feeder link structure is not perfectly calibrated, the system achievable rates decrease even though precoding is used. Thus, the multigateway feeder link hardware architecture shall be carefully designed in order to preclude the possible interference effect. In any case, as long as precoding is employed, the inter-feeder link interference can be reverted at the transmit side but; however, certain performance loss occurs with respect to the ideal interference-free feeder link system. 

\subsection{CSI Feedback Errors}

Precoding severally relies on the CSI integrity. Preserving the quality of the estimation carried out by the receiver requires an ideal feedback mechanism which is impossible to implement in real systems. However, as it happens in terrestrial communications, broadband satellite standards are including CSI feedback mechanisms for supporting precoding techniques. This is the case of DVB-S2X \cite{ETSI} where for the first time certain feedback technique is offered to the system designer. In the following this mechanism is briefly described.

The measurement and quantizing  process of CSI feed back in each gateway is assumed to be continuous and to be reported on the return channels through a signalling table only when significant changes are detected. The maximum delay required for estimation and delivery to the gateway via the interaction channel shall be no more than 500 ms \cite[Annex E.4]{ETSI}, but this delay should be minimized to maximize capacity gain.

Each user shall estimate and report the channel transfer function to the gateway as a set of complex-valued coefficients. These coefficients should be estimated by a set of 32 orthogonal Wash-Hadamard sequences plus 4 padding symbols. In this context, every feed signal should incorporate a different sequence so that the receiver is able to estimate the channel effect of 31 interfering feed signals. With this, the overall channel matrix cannot be estimated but; however, the closest 31 feed signals are the ones whose largest interference power levels so that the rest can be ignored for precoding purposes. Note that this also reduces the inter-gateway communication overhead.

Under this context, each user can feed back a maximum 7 digits (i.e. maximum 3 digits before decimal point and 4 digits after decimal point)  for both amplitude and phase of each channel element  such that
\begin{equation}
\left(\mathbf{H} \right)_{i,j}=ddd.dddd^{\angle aaa.aaaa},
\end{equation}
where it is evident that there are 7 bits for the magnitude so as 7 for the phase. Remarkably, this CSI report considers the effects of not only the user channel but also the tentative feeder-link imperfections. This is considered in the numerical results section.

\section{Simulation Results}

\subsection{Simulation Setup}

In order to illustrate the performance of our proposal, this section presents the simulation results related to the considered scenario in the previous sections. The simulation setup is based on an array fed reflector antenna whose channel gain matrix has been provided by ESA in the framework of a study on next generation multibeam  satellite systems. The number of feeds and beams  is assumed to be $N = 155$ and $K=100$, respectively, which is covering the whole Europe area such that $N \geq K$.
\begin{table}[h!]
\caption{USER LINK SIMULATION PARAMETERS} 
\begin{center} 
\begin{tabular}{ | l | l |}
\hline
\begin{math}\mathbf{Parameter}\end{math}&\begin{math}\mathbf{Value}\end{math} \\ \hline
    Satellite height & 35786 km (geostationary) \\ \hline
   Satellite longitude, latitude &\begin{math} 10^\circ East, 0^\circ\end{math} \\ \hline
  Earth radius& 6378.137 Km   \\ \hline
Feed radiation pattern & Provided by\begin{math} ESA\end{math} \\ \hline
    Number of feeds N & 245  \\ \hline
User location distribution&Uniformly distributed \\ \hline
Carrier frequency&20 GHz (Ka band) \\ \hline
Atmospheric fading&Just rain fading \\ \hline
Roll-off factor&0.25 \\ \hline
User antenna gain&41.7 dBi \\ \hline
 clear sky gain&17.68 dB/K \\ \hline
Frequency&20$\times 10^9$ \\ \hline
        \end{tabular}
     \vspace{-0.4cm}  \label{tab:label}
\end{center}
\end{table}
\begin{table}[h!]
\caption{DVB-S2x MODCOD parameters }
\begin{center}
  
	\begin{tabular}{| c | c | c |}
    \hline
    \textbf{ ModCod} &\textbf{Efficiency}  & \textbf{Required
   SINR [dB]}\\
 \textbf{ mode} & \textbf{Info bit / symbol}  & \textbf{(with approx. impl. losses)}\\
	\hline  QPSK\_2/9 &    0.434  &   -2.85 \\ 
	\hline QPSK\_13/45 &    0.567& -2.03\\
  \hline QPSK\_9/20 &0.889 & 0.22 \\
  \hline QPSK\_11/20 &    1.088  &   1.45 \\ 
	\hline 8APSK\_5/9-L& 1.647& 4.73 \\
	\hline 8APSK\_26/45-L &1.713 &5.13 \\ 
	\hline 8PSK\_23/36 & 1.896& 6.12 \\ 
	\hline 8PSK\_25/36 &2.062& 7.02 \\
	\hline 8PSK\_13/18 &2.145 &7.49 \\ 
	\hline 16APSK\_1/2-L   & 1.972&5.97 \\ 
	\hline 16APSK\_8/15-L &2.104 &6.55 \\
	\hline 16APSK\_5/9-L& 2.193 &6.84 \\
	\hline 16APSK\_26/45 &2.281 &7.51 \\ 
	\hline 16APSK\_3/5 &2.370 &7.80 \\ 
	\hline 16APSK\_3/5-L &2.370 &7.41 \\
	\hline 16APSK\_28/45 &2.458 &8.10 \\
	\hline 16APSK\_23/36 &2.524 &8.38 \\ 
	\hline 16APSK\_2/3-L &2.635 &8.43 \\
	\hline 16APSK\_25/36 &2.745 &9.27 \\
	\hline 16APSK\_13/18 &2.856 &9.71 \\ 
	\hline 16APSK\_7/9 &3.077 &10.65\\
	\hline 16APSK\_77/90 &3.386 &11.99 \\ 
	\hline 32APSK\_2/3-L& 3.289 &11.10\\ 
	\hline 32APSK\_32/45& 3.510 &11.75 \\ 
	\hline 32APSK\_7/9 &3.841 &13.05\\
	\hline 64APSK\_32/45-L &4.206 &13.98 \\
	\hline 64APSK\_11/15 &4.338 &14.81 \\
	\hline 64APSK\_7/9 &4.603 &15.47 \\
	\hline 64APSK\_4/5 &4.735 &15.87 \\
	\hline 64APSK\_5/6 &4.933 &16.55 \\
	\hline 128APSK\_3/4 &5.163 &17.73 \\
    \hline
  \end{tabular}
\vspace{-0.4cm} \label{tablemodcod}
\end{center}
\end{table}

We assume that the users are uniformly distributed over the coverage region and there is only one user per beam. Results have been averaged for a total of 500 user link channel realizations. Note that only atmospheric fading due to rain is considered in the user link channel and further refinements of the channel  are not considered. Furthermore, the randomness of the channel is due to the user positions which are assumed to be uniformly distributed within the beams. Recall that, full frequency reuse among beams have been considered in this contribution. Table \ref{tab:label} represents the detail of simulation parameters.

We compute the SINR for each user, after interference mitigation and then its throughput (bit/s) is inferred according to  DVB-S2X standard for a packet error rate (PER) of $10^{-6}$ \cite{ETSI}. Table \ref{tablemodcod}  provides a one-to-one relationship between the required received SINR and the efficiency (bits/symbol) that are achieved by the different adaptive modulation and coding modes included in the DVB-S2X standard.

For a best practice and in order to clarify the performance of proposed multiple gateway architecture, we consider the upper-bound of the system achievable rates that result from considering a single gateway system architecture in \cite{miguel11}. 

\subsection{Ideal Multigateway System Architecture}

\begin{figure}[h!]
\centering
\includegraphics[scale=0.3]{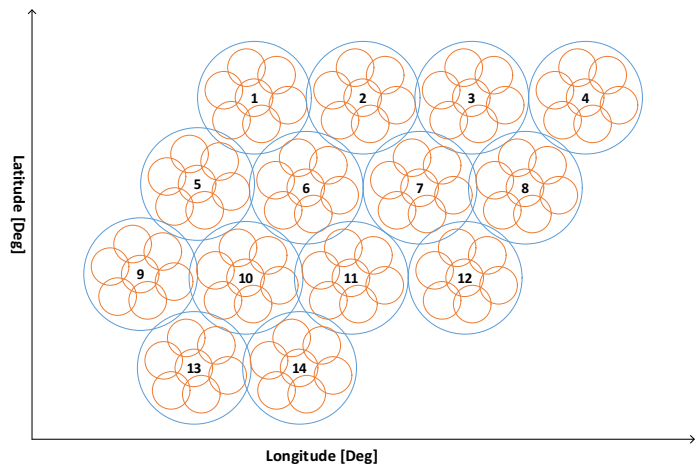}
\vspace{-0.4cm} \caption{ A set of 14 cluster composed by 7 beams is depicted in the figure. This will be the reference architecture assuming that all beams operate in the same frequency band. Remarkably, the precoder not only has to mitigate the inter-cluster interference but the intra-cluster one. }
 \label{clustering}
\end{figure}

\begin{figure}[h!]
\centering
\includegraphics[scale=0.3]{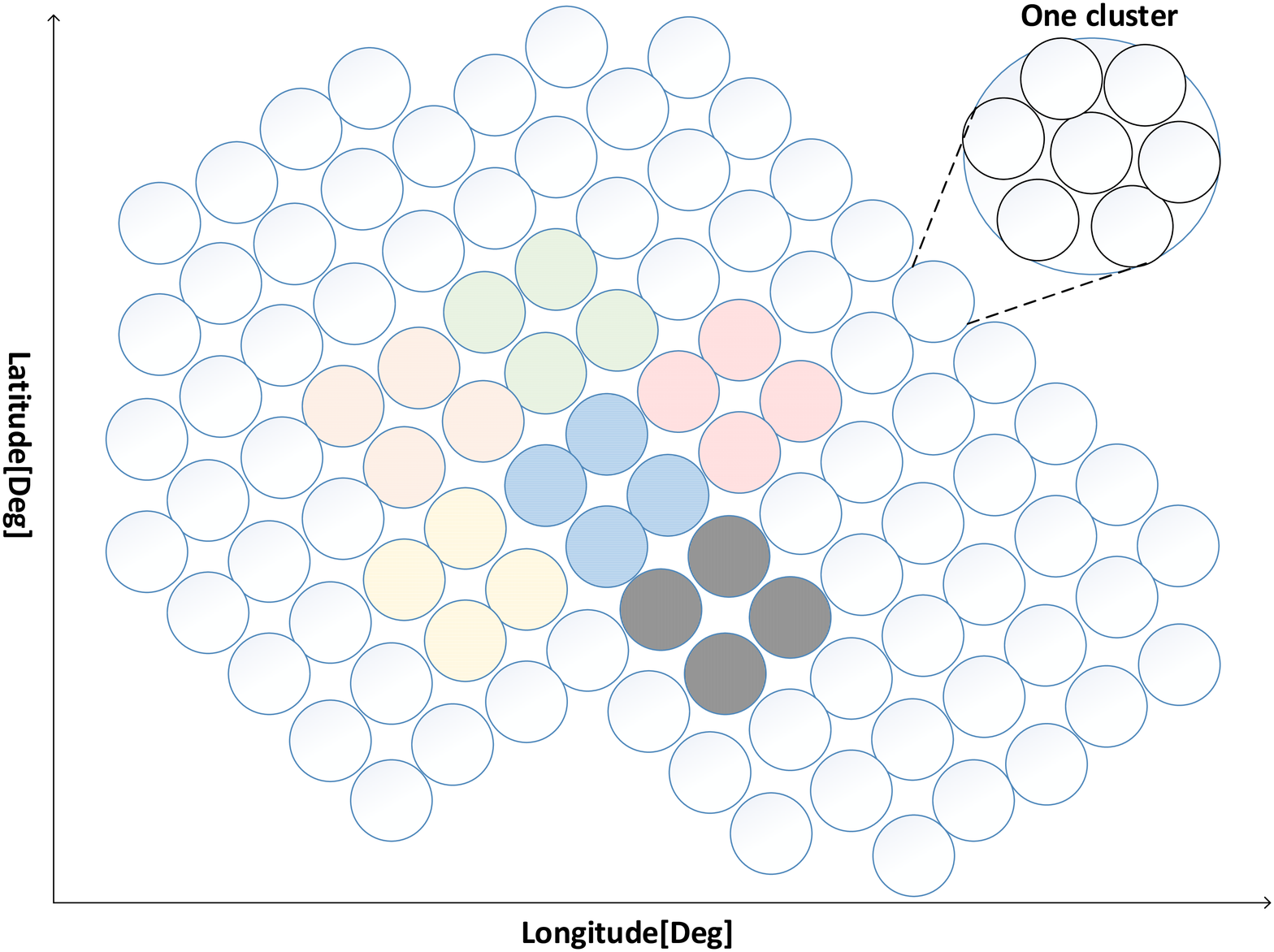}
\vspace{-0.4cm} \caption{ Cooperation architecture between 4 adjacent clusters: it is assumed that each gateway can cooperate with only 3 gateways whose beams are adjacent to them. The cooperative clusters are depicted with the same colour. }
 \label{fig:m_GW8}
\end{figure}

This section presents the simulation results related to the scenarios described in Section II and III. The user link in the downlink is assumed to operate at 20 GHz Ka-band.  A total of $G=14$ gateways are considered so that each gateway is serving a cluster of 7 or 8 beams. Fig. \ref{clustering} depicts the overall system architecture.  It can be observed that the $g$-th gateway is serving a set of 7 beams $K_g = 7$. Several collaborative schemes are presented in the following for the sake of completeness:
\begin{itemize}
\item $\textbf{Scenario 1}:$ The individual cluster multibeam processing without gateway coordination so that each gateway processes its beams independently. In this context, the ZF precoding in $m$-th gateway can be expressed as
\begin{equation}
\mathbf{T}_{g}=\beta_{\text{ICM}}\mathbf{H}_{g}^{H}(\mathbf{H}_{g}\mathbf{H}_{g}^{H})^{-1},
\label{simple}
\end{equation} 
where $\beta_{\text{ICM}}$ is set to preserve the gateway power constraint assumed to be $\frac{P}{G}$ for all of them. This is referred to  Individual Cluster Multibeam processing (ICM).

\item $\textbf{Scenario 2}:$ 4 gateways interchange their CSI so that each gateway only has access to 3 interfering matrices. This is referred to  4 Gateways  Collaboration (4GC). See Fig. \ref{fig:m_GW8}.

\item $\textbf{Scenario 3}:$ 7 gateways interchange their CSI so that each gateway only has access to 3 interfering matrices. This is referred to  7 Gateways  Collaboration (7GC).

\item $\textbf{Scenario 4}:$ Gateway $g$ (respectively for all the gateways) collaborates with all gateways.  This is referred to  Gateway Collaboration Multibeam processing (GCM).

\item $\textbf{Scenario 5}:$ Single gateway scenario where all data and CSI is located at the same transmitter. This scenario is the Reference scenario (Ref). 

\item $\textbf{Scenario 6}:$ Gateway $g$ (respectively for all the gateways) collaborates with all the gateways that serve clusters that are directly adjacent to cluster $g$ by means of transmitting the rank one approximation of their channels as we discussed in section IV A.  This is referred to  Limited Multi-gateway Collaboration processing (LMC).
\end{itemize}

Table III describes the communication overhead associated to each cooperative scheme. As we anticipated in the previous section, the LMC scheme offers a large reduction of the communication overhead.

\begin{table}[h!]
\caption{Cooperation Overhead Comparison} 
\begin{center} 
\begin{tabular}{ | l | l |}
\hline
Cooperation Scheme & Total Number of Complex Numbers to be Shared \\ \hline
4GC & 57288 \\ \hline
7GC & 100254 \\ \hline
GMC & 200508   \\ \hline
LMC & 143 \\ \hline
   
\end{tabular}
\vspace{-0.4cm}  \label{tab:label}
\end{center}
\end{table}


\begin{figure}[h!]
\centering
\includegraphics[scale=0.6]{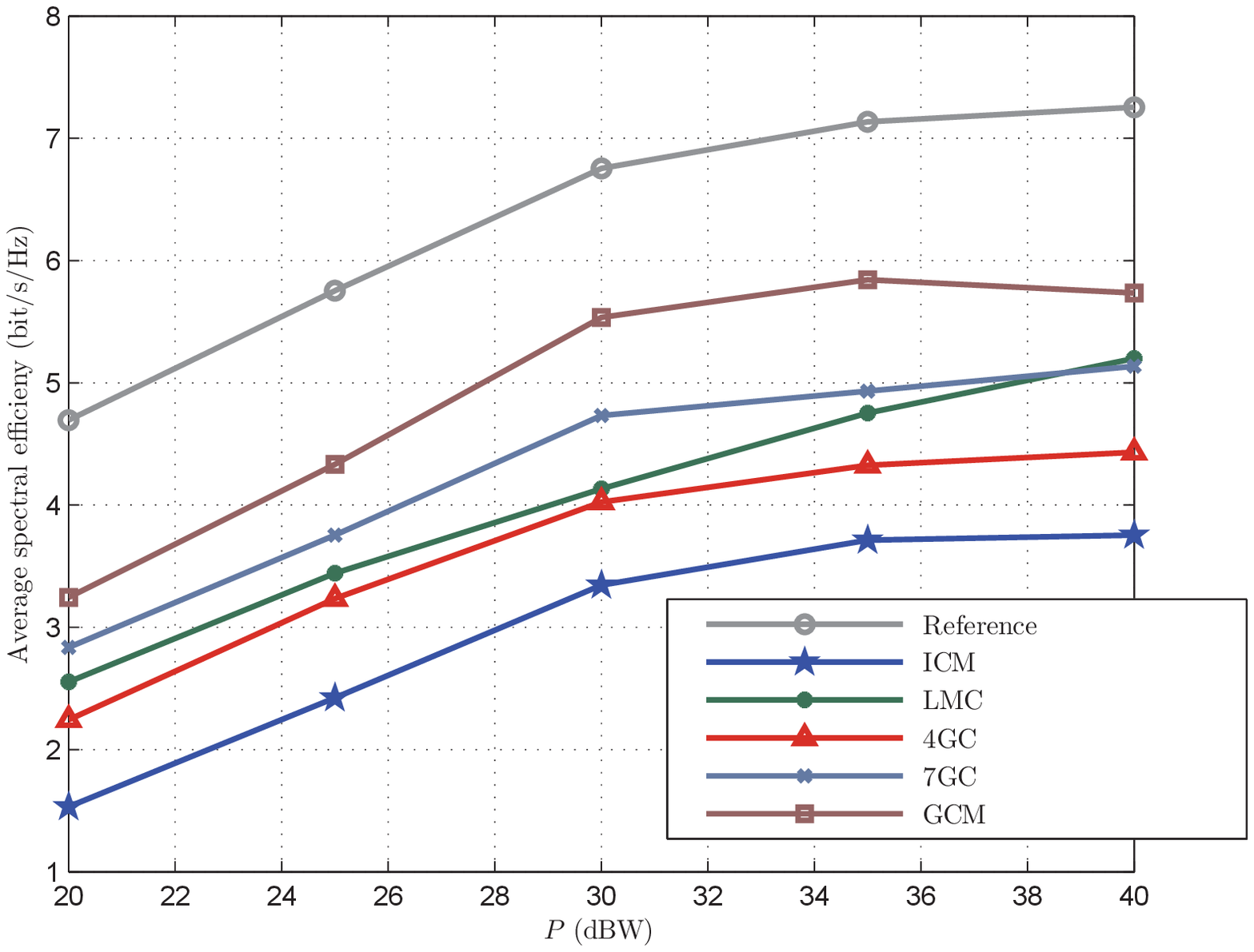}
 \caption{Spectral efficiency considering multigateway block regularized precoding and different collaborative architectures. The intra-cluster interference is mitigated via MMSE precoding} 
 \label{Figmmse}
\end{figure}

\begin{figure}[h!]
\centering
\includegraphics[scale=0.6]{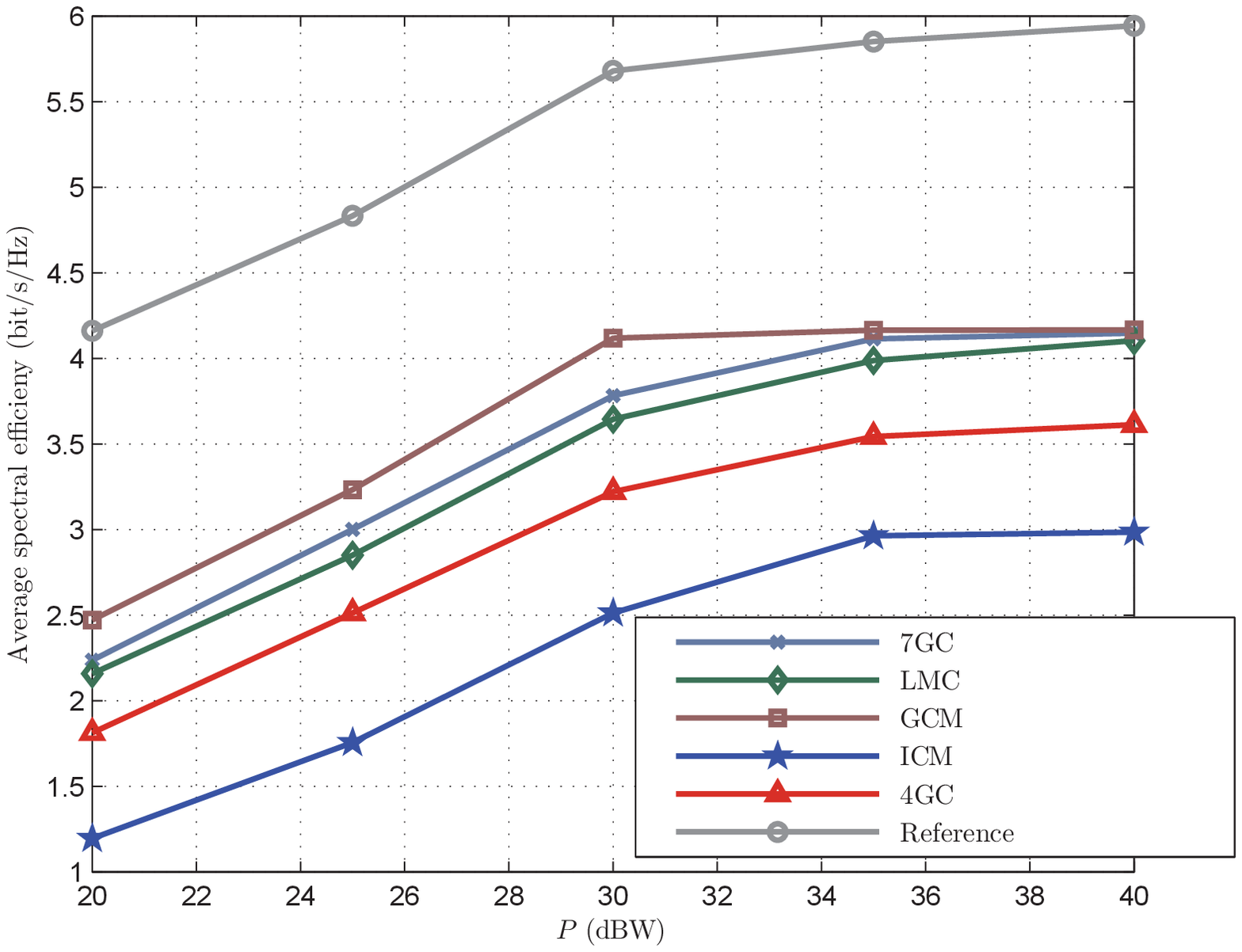}
 \caption{Spectral efficiency considering multigateway block regularized precoding and different collaborative architectures. The intra-cluster interference is mitigated via ZF precoding}
 \label{Figzf}
\end{figure}

Figure \ref{Figmmse} and \ref{Figzf} present the spectral efficiency when either MMSE or ZF is used as precoding matrix for mitigating the intra-cluster interference respectively. For both cases it is shown that the ICM scheme has the lowest achievable rate due to lack of interference mitigation  among clusters. 4GC and 7GC achieve reasonable performance and  the proposed multiple gateway scheme with CSI sharing  among adjacent clusters (i.e. GCM) shows better performance.  Consequently, the higher the coordination among gateways is, the higher the achievable rates are . Note that MMSE delivers higher spectral efficiencies than ZF.

Note that the proposed LMC offers a good trade-off between gateway cooperation overhead and overall system performance.

\subsection{Non-Ideal Feeder Link System Architecture}

\begin{figure}[h!]
\centering
\includegraphics[scale=0.6]{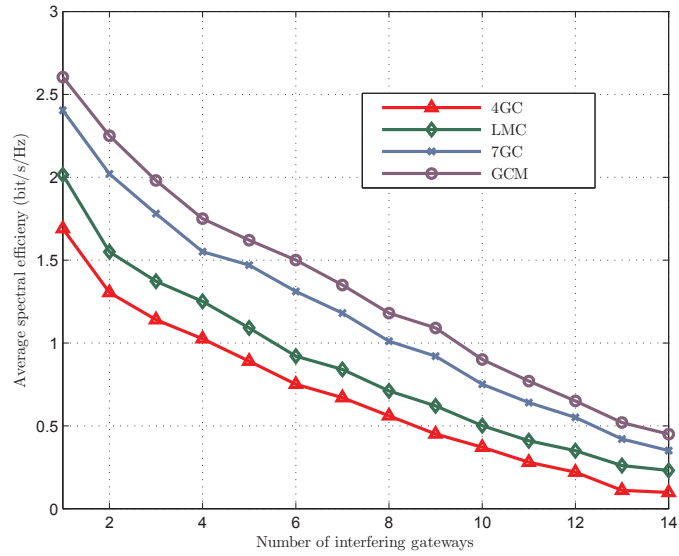}
 \caption{Spectral efficiency versus number of interfering gateways. The interfering parameter is set to 1 ($\rho =1$). The feeder links uncalibrations decrease the system capacity severally.}
 \label{Figrho}
\end{figure}

In this section we  analyse  the scenario presented in the section IV. We consider $\rho = 1$, which is a worst-case scenario. Again, simulation results use the average total throughput as performance measurement. Figure \ref{Figrho} compares the results related to all scenarios described above considering that for each gateway receives interference from 1 to 14 gateways. The transmit power is set to $P=30$ dBW and MMSE precoding is used for mitigating the intra-cluster interference.

From Figure \ref{Figrho} it is evident the dramatical effect of feeder links mismatches in multi-gateway multibeam architecture even though precoding is performed. This effect appears for any cooperative architecture. It is important to remark that even if only one interfering gateway is considered, the average spectral efficiency decreases up to the 54\% with respect to the ideal feeder-link scenario.

\subsection{Limited Feedback}

Let us consider the case where the CSI is not perfect but it is obtained via DVB-S2X feedback mechanisms as we described in section IV B.

\begin{figure}[h!]
\centering
\includegraphics[scale=0.6]{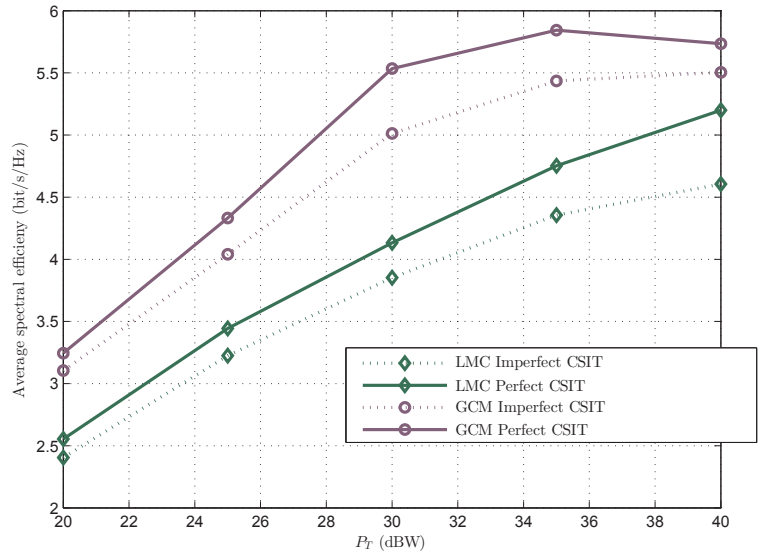}
 \caption{Spectral efficiency versus transmit power considering GMC and LMC processing with CSI errors. The performance decrease is high whenever the available CSI is not perfect.}
 \label{Figcsit}
\end{figure}

Figure \ref{Figcsit} depicts the performance degradation whenever quantized feedback is used instead of the perfect one. Both for the LMC and GMC case, the spectral efficiency decrease is notorious.

\section{Conclusion} 
 This paper presents a precoding design for multigateway multibeam satellite systems. These architectures suffer from a large intra and inter cluster interference. The proposed technique mitigate both interference types, leading to large spectral efficiencies. Although the presented method assumes that gateways share the CSI (i.e. all channel matrix), limited cooperative schemes are introduced. In addition, whenever certain feeder link imperfections are considered, it is shown that the SMSE and thus the achievable rates decrease. The proposed techniques were validated through the DVB-S2X standard which recently proposed certain CSI feedback mechanisms.

\appendices 
\section{Proof of Theorem 1}

The SMSEs can be rewritten so that
\begin{equation}
\text{SMSE}_{\text{interference}} = \sum_{i=1}^{K}\frac{1}{\frac{G}{P}+\lambda_{i}(\mathbf{H}_{u}\mathbf{H}_{f}\mathbf{H}_f^H\mathbf{H}_u^H)},
\end{equation}
\begin{equation}
\text{SMSE}_{\text{no-interference}} = \sum_{i=1}^{K}\frac{1}{\frac{G}{P}+\lambda_{i}(\mathbf{H}_{u}\mathbf{H}_u^H)}.
\end{equation}

Under this context and manipulating the previous equations, the theorem 1 holds as long as 
\begin{equation}
\lambda_{i}(\mathbf{H}_{f}\mathbf{H}_{f}^H\mathbf{H}_u\mathbf{H}_u^H) \leq \lambda_{i}(\mathbf{H}_{u}^H\mathbf{H}_u),
\label{ineq}
\end{equation}
for $i=1, \ldots, K$. Bearing in mind that,
\begin{equation}
\lambda{i}(\textbf{H}_{u}\textbf{H}_{f}\textbf{H}_f^H\textbf{H}_u^f)=\sigma_{i}^{2}(\textbf{H}_u\textbf{H}_{f}\textbf{H}_{f}^H\textbf{H}_{u}^H),
\end{equation}
\begin{equation}
\lambda{i}(\textbf{H}_{u}\textbf{H}_{u}^H)=\sigma_{i}^{2}(\textbf{H}_{u}\textbf{H}_{u}^H),
\label{hasbe}
\end{equation}
and considering that $\mathbf{H}_f$ has the following SVD decomposition $\mathbf{H}_{f}=\mathbf{U}_{f}\boldsymbol{\Sigma}_{f} \mathbf{V}^{H}_{f}$, we have that
\begin{equation}
\mathbf{H}_f\mathbf{H}_f^H = \mathbf{U}_{f}\boldsymbol{\Sigma}_{f}\boldsymbol{\Sigma}_{f}^H \mathbf{V}^{H}_{f}.
\end{equation}
Writing $\boldsymbol{\Sigma}_{f}\boldsymbol{\Sigma}_{f}^H = \mathbf{S}_f$ it is easy to observe
\begin{equation}\label{65211}
\textbf{S}_{f}= \left(
\begin{array}{cc}
 \mathbf{z}_{N_{g}\times N_{g}} &\textbf{0}_{ N\times(N-N_{g})}\\
  ~~~~~~\textbf{0}_{(N-N_{g})\times N_{g}} &~~~~~~\textbf{0}_{(N-N_{g})\times (N-N_{g})}
\end{array}\right),
\end{equation}
where $(\boldsymbol{\Sigma}_{f}\boldsymbol{\Sigma}^{H}_{f})$ has only $N_{g}$ non-zero  singular values (i.e. $\mathbf{z}$),  as $\textbf{S}_{g}$ has rank equal to $N_{m}$.

Since $\mathbf{U}$ is a unitary matrix, for any matrix (same as $\mathbf{H}_u^{H}$ in this study) it holds that
\begin{equation}
 \sigma_{i}\left(\mathbf{U}\mathbf{H}_{u}^{H}\mathbf{H}_{u}\right)=\sigma_{i}\left(\mathbf{H}_{u}^{H}\mathbf{H}_u\right).
\label{FGH123}
\end{equation}
Then, the right hand side of \eqref{ineq} can be worked out as
\begin{equation}
\sigma_{i}\left(\mathbf{U}_f\mathbf{S}_f\mathbf{U}_f^{H}\mathbf{H}_{u}^{H}\mathbf{H}_{u}\right)=\sigma_{i}\left(\mathbf{S}_{f}\mathbf{U}^{H}_{f}\mathbf{H}_{u}^{H}\mathbf{H}_{u}\right).
\end{equation}
By the following definition $\mathbf{G}_{f}\triangleq \mathbf{U}_f^{H} \mathbf{H}_{u}^{H}\textbf{H}_{u}$ and considering \eqref{FGH123}, proving \eqref{ineq} is equivalent to checking
\begin{equation}
\sigma_{i}(\mathbf{G}_{f}) \geq \sigma_{i}(\mathbf{S}_{f}\mathbf{G}_{f}),
\end{equation}
for $i=1, \ldots , N_g$.
Now, remind that  $\mathbf{G}_f$ is of size $N \times N$ as follows
\begin{equation}
\mathbf{G}_{f}={\mathbf{G}_{1} \choose \mathbf{G}_{2}}
\end{equation}
where  both sub-matrices of $\mathbf{G}_{f}$ are of size $N_{g}\times N$.  Then, we have that
\begin{equation}
\mathbf{S}_{f}\mathbf{G}_{f}={\mathbf{G}_{1} \choose \mathbf{0}}
\end{equation}
It is clear that \\

\begin{equation}\label{charba}
\mathbf{G}^{H}_{g}= \left(\begin{array}{cc}
 \small\mathbf{G}^{H}_{1} &\mathbf{G}^{H}_{2}\\
    ~~~~~~0 &~~~~~~0
\end{array}\right),
\end{equation}
where $\mathbf{G}^{H}$ is matrix of size $N \times N$ whose singular values are
\begin{equation}
\boldsymbol{\sigma}( \mathbf{G}^{H}_{f} )= \left( \sigma_{1}(\mathbf{G}_{f}), \sigma_{2}(\mathbf{G}_{f}), \ldots, \sigma_{N_{g}}(\mathbf{G}_{f}), \mathbf{0}_{N - N_g} \right),
\label{WER1122}
\end{equation}
where $\mathbf{0}_{N - N_g}$ is a vector of dimension $N-N_g$ whose entries are equal to zero. By the interlacing property in Lemma \ref{LEMMASD} tells that
\begin{equation}
\sigma_{i}(\mathbf{G}_{f}) \geq \sigma_{i}(\mathbf{S}_{f}\mathbf{G}_{f}).
\end{equation}

\bibliographystyle{IEEEtran}
\bibliography{references}

\end{document}